# PHENOMENOLOGY OF STANDARD MODEL IN SPONTANEOUSLY BROKEN MIRROR SYMMETRY

*Igor T. Dyatlov \**
*Scientific Research Center "Kurchatov Institute"*
*Petersburg Institute of Nuclear Physics, Gatchina, Moscow, Russia*

Violated mirror symmetry (MS) is capable of reproducing observed qualitative properties of weak mixing for quarks and leptons. In violated MS, lepton phenomenology—that is, small neutrino masses and mixing properties different from those of quarks—requires the Dirac nature of neutrinos and existence of processes that change the total lepton number. Such processes involve heavy mirror neutrinos, and therefore occur at very high energies. $CP$ non-conservation would mean here that the parity conserving MS Lagrangian must be non-invariant to both time reversal $T$ and (according to the $CPT$-Theorem) the charge conjugation $C$. All these properties create appropriate conditions for leptogenesis, a mechanism for generating baryon-lepton asymmetry of the Universe in violated MS models.

## 1. Introduction

The main objective of the present and previous papers by the author [1-3] is to search for a mechanism that would be capable of reproducing observed qualitative structure of weak mixing matrices (WMM) for quarks and leptons. Unlike other studies of fermion spectrum properties in the Standard Model (SM) [4-7], we are specifically interested in the ability to reproduce the WMM structure. Another, not less important property of the spectra—the observed hierarchy of quarks and charged lepton masses—is considered as data given by experiment.

In [1-3], mass hierarchy and involvement of so-called mirror states [8] are shown to produce the known hierarchy of quark WMM elements (the Cabibbo-Kobayashi-Maskawa (CKM) matrix, the Wolfenstein parametrization [10]) and explain the appearance of a lepton WMM with different properties. For this purpose, the SM fermion system should be supplemented by massive mirror analogs, their only difference being weak interactions (i.e., substitution of left-handed weak currents for right-handed ones) and masses.

The proposed scenario can be described as follows. The total initial fermion system includes three generations of weak isodoublets and three generations of weak isosinglets of the $SU$(2) flavor symmetry, $f = \bar{u}, \bar{d}$. The respective operators are:

$$\Psi_{LR} = \psi_L + \Psi_R, \left(T_W = \frac{1}{2}\right), \quad \Psi_{RL} = \psi_R + \Psi_L, (T_W = 0). \tag{1}$$

Here, $T_W$ is the weak isospin, $L$ and $R$ are the left- and right-handed chiralities of massive Dirac quarks and leptons $\Psi_{LR}$ and $\Psi_{RL}$.

\* E-mail: dyatlov@thd.pnpi.spb.ru

SM particles ($\psi_L$, $\psi_R$) appear upon breaking of the obvious "mirror" symmetry (MS):

$$\Psi_R \leftrightarrow \psi_L, \quad \Psi_L \leftrightarrow \psi_R, \qquad (2)$$

when only those states defined by the chiral parts ($\Psi_R$, $\Psi_L$) spontaneously acquire heavy masses in addition to the masses $\Psi_{LR}$ and $\Psi_{RL}$. The $\Psi_R$, $\Psi_L$ masses must be much heavier than even the heaviest fermions of SM. This very property distinguishes our proposed mirror system from systems earlier described by other authors [8,11,12] and is the most important condition for the formation of observed WMM structures. Previous mirror symmetry scenarios were designed to explain only the paradoxes of SM parity non-conservation, and the new particles did not have to be very heavy.

The large mirror masses, mass hierarchy of charged fermions, and weak *SU*(2) flavor symmetry appear to be the major factors responsible for the appearance of the observed properties in both quark and lepton WMMs. At that, neither introduction of additional couplings nor fitting of constants is required.

For leptons, this results (apart from the observed WMM properties) in the inevitable appearance of SM neutrino masses that are extremely small compared to the $\bar{u}$ (up) and $\bar{d}$ (down) masses of charged fermion families. To reproduce these properties, SM neutrinos must be of the Dirac type and have an inverse mass spectrum. The smallness of neutrino masses is an accompanying element of choice of suitable WMM properties. This scenario is close to the widely known see-saw mechanism [7,13]. Mirror neutrinos are much heavier than charged mirror leptons, although they may be considerably lighter than the heavy Majorana particles in the see-saw scenario ($M_{mir}^{(v)} \ll M_{gut}$).

The principles of weak *SU*(2) symmetry violation mechanism in SM [14] are used in [3] to construct a model of spontaneous MS breaking (1)-(2). In the present paper, we discuss general phenomena, i.e., the implications of MS breaking. Specific effects (Sections 6 and 7) are too complicated for quantitative consideration due to the unavoidable in this scenario non-perturbative Yukawa coupling of heavy mirror fermions with the real SM Higgs boson [15]. The strong coupling is a compulsory element accompanying incorporation of *SU*(2) symmetry breaking into the model that is a close analog of SM (Section 5).

A common property of spontaneous MS violation models is non-invariance under time reversal $T$. This non-invariance is independent of the MS violation mechanism and should already be present in the symmetric Lagrangian. The appearance of this property has purely phenomenological reason; similar to SM, it is necessary to introduce complexity to account for the observed $CP$-violation. In MS systems, however, the spatial parity $P$ is preserved by the very definition of mirror symmetry.[1] Weak currents (see Eq.(3)) have a vector nature; MS violation results in simultaneous non-conservation of $P$ and charge conjugation $C$, while generally preserving $CP$-symmetry. To reproduce here the typical $CP$ mechanism of SM—that is, the complexity of quark and lepton WMMs—it is necessary to incorporate terms non-invariant to time reversal $T$ into the MS-Lagrangian. According to the Lüders–Pauli $CPT$-Theorem [8], such terms (with $P$ preserved) should also violate $C$, but they represent $T$-violation since complexities are specifically typical of $T$ non-invariance.

Another general consequence of MS breaking concerns heavy Dirac mirror leptons at very high energies. In weak processes involving mirror particles, a 100 percent change in the total lepton number may take place. This effect is again associated with phenomenology; it is present in the scenario where light neutrinos acquire extremely small masses and WMM different from the CKM quark matrix. Also, as mentioned earlier, all neutrinos must be Dirac here.

The above two properties can form the basis for the MS-scenario of leptogenesis that, under certain conditions [17], produces the baryon-lepton asymmetry of the Universe [16].

The MS violation scenario also involves generation of new heavy fermions and scalar bosons. It postulates the presence of the second Higgs field and bosons with the charges (0,+;-,0), similar to the case of the $K - \bar{K}$-meson doublets. Their masses are unknown. Non-perturbative couplings, inevitable in MS for these bosons, do not permit quantitative evaluations.

Neutral processes with generation changes [18] are of high importance. Non-perturbative couplings hinder evaluations here as well. Probabilities now depend not only on the general principles of the MS-scenario but also on the details of Yukawa coupling matrices. In our model, they are definitely not large since they are suppressed by the orders of the ratios $m_{\text{SM}}/M_{mir}$, where $m_{\text{SM}}$ is the mass of SM particles and $M_{mir}$ is the mass of mirror states. In principal orders, defined by the general properties of MS breaking, neutral transitions between generations are absent.

The non-perturbative character of Higgs couplings in the MS-scenario, which impedes quantitative evaluations, does not have any perceivable impact on the properties in question. This non-perturbative character may be responsible for the Lagrangian parameters being different from the physical masses of particles, while not, in general, changing the WMM properties defined by the diagonalization of these initial parameters.

---

[1] The presence of the QCD strong CP violation terms ($\Theta F \tilde{F}$) also contradicts MS principles.

In Section 2, we discuss those results in [1-3] that can be used to define the $CP$- and $T$-properties of the MS-scenario. Sections 3 and 4 consider the general character of discrete symmetry violations in quark and lepton WMMs. Section 5 describes other possible manifestations of spontaneous MS violations. One of such manifestations—the violation of the total lepton number for heavy mirror states—is discussed in Section 6. In Section 7, Conclusions, we summarize our discussion of MS implications resulting from the model in question. The Appendix explains the mechanism for the approximate calculation of the mass matrix (16).

## 2. Models of Mirror-Symmetric Lagrangian

This section discusses some results of [1-3], modified to account for the specific conditions under which $CP$-violating factors appear in the MS-scenario. Fermion states in the MS-Lagrangian must be expressed exclusively in terms of the operators (1), $\Psi_{LR}$ and $\Psi_{RL}$, with all standard invariances taken into consideration. At that, the kinetic terms and gauge couplings will automatically fall into parts dependent only on the components $\psi$ or $\Psi$. Simultaneously, they will also group by the chiralities $R$ and $L$ and the flavors up ($\bar{u}$) and down ($\bar{d}$) (with the exception of the weak interaction).

We obtain the sum of the MS-Lagrangians for $\varphi$ and $\Psi$, with left-handed weak current for $\psi_L$ and right-handed weak current for $\Psi_R$:

$$\mathcal{L}_W = g\bar{\Psi}_{LR}\gamma^\mu \tau^A \Psi_{LR} W_\mu^A = g\{\bar{\psi}_L \gamma^\mu \tau^A \psi_L + \bar{\Psi}_R \gamma^\mu \tau^A \Psi_R\} W_\mu^A. \tag{3}$$

Similar to SM, all these parts are diagonal in terms of generation indices (assuming summation over them).

The isodoublets $\Psi_{LR}$ and isosinglets $\Psi_{RL}$ define massive Dirac fermions. Mass terms of the $SU$(2)-invariant Lagrangian have the form:

$$\mathcal{L}_M = A_a\left(\bar{\Psi}_{LR}^{(f)} \Psi_{LR}^{(f)}\right)_a + B_b^{(f)}\left(\bar{\Psi}_{RL}^{(f)} \Psi_{RL}^{(f)}\right)_b, \tag{4}$$

where $a, b$ = 1, 2, 3 are generation indices and $A$ does not depend on the flavor $f = (\bar{u},\bar{d})$. Eq.(4) appears to produce transitions between the chiral parts $\Psi_R$, $\Psi_L$ and $\psi_L$, $\psi_R$:

$$\mathcal{L}_M = A_a\left(\bar{\Psi}_R \psi_L + \bar{\psi}_L \Psi_R\right)_a^{(f)} + B_b^{(f)}\left(\bar{\Psi}_L \psi_R + \bar{\psi}_R \Psi_L\right)_b^{(f)}. \tag{5}$$

The operators $\Psi_{LR}$ and $\Psi_{RL}$ can be transformed in generation space with unitary matrices $U$. At that, the diagonal parts of the Lagrangian with the coupling constants independent of indices do not change, while the masses $A$ and $B$ in Eq.(4) and Eq.(5) become mass matrices. Only those transformations are acceptable that do not violate $SU$(2)-invariance (4) and (5), i.e., the flavor independence of the factor $A$, which defines mass properties of isospinors:

$$A^{(u)} = A^{(d)}. \tag{6}$$

Eq.(6) should be satisfied for both the diagonal masses (4) and mass matrix elements when changing to a different direction $U\Psi$ in generation space. This condition is dictated not only by *SU*(2)-invariance but also by phenomenology. As shown in [1], condition (6) is compulsory (along with quark mass hierarchy) for the reproduction of all observed qualitative properties of WMM, the CKM matrix for quarks [9].

The spontaneous MS breaking must change the invariant system into one of two equivalent states. In one state, there are heavy $\Psi$ and the left-handed current (3) of light $\psi$ particles. This is the MS generalization of SM. In the second state, heavy $\psi$ accompany the right-handed weak current of light $\Psi$. All other properties of these two states—masses, interactions, mixings—have to remain identical in order not to distinguish their *R* and *L* properties. Such are the conditions of mirror symmetry. In this regard, the existence of two scalars of different parity appears to be necessary.

Suitable properties become available in the spontaneous breaking model [3], in which Yukawa couplings with isodoublet bosons $\varphi_1$ (scalar) and $\varphi_2$ (pseudoscalar) are present[2]:

$$\begin{aligned}\mathcal{L}_\varphi &= \bar{\Psi}_{LR}h^{(f)}\Psi^{(f)}_{RL}\varphi_1 + \bar{\Psi}_{LR}h^{(f)}\gamma_5\Psi^{(f)}_{RL}\varphi_2 + \text{c. c.} = \\ &= \bar{\Psi}_R h^{(f)}\Psi^{(f)}_L \phi_1 + \bar{\psi}_L h^{(f)}\psi^{(f)}_R \phi_2 + \text{c. c.} ;\end{aligned} \tag{7}$$

$$\phi_1 = \varphi_1 - \varphi_2, \qquad \phi_2 = \varphi_1 + \varphi_2. \tag{8}$$

The coupling constants *h* (matrices or vectors in generation space) are the same for interactions with $\phi_1$ and $\phi_2$. Only then, after spontaneous breaking, all properties of arising systems (except weak systems) become identical. Moreover, without this identity, the proposed mechanism of appearance of observed CKM matrix properties is practically impossible.

Without MS breaking, the spatial parity *P* is preserved. This is why MS is being introduced. Removing MS (the difference of *L*, *R*-weak currents) results in non-conservation of both *P* and *C*. It would be natural to consider MS violations as the only cause of non-conservation of all discrete symmetries; any other sources are deemed to be redundant. Then, prior to violation, the MS-Lagrangian would be invariant to both *P*- and *C*-transformations, and this means that all possible Yukawa coupling-matrices are real-valued, because:

$$\bar{\Psi}^{(c)}_{1a} h_{ab} \Psi^{(c)}_{2b} = -\Psi^T_{1a} C^+ h_{ab} C \bar{\Psi}^T_{2b} = \bar{\Psi}_{2b} h_{ab} \Psi_{1a},$$

$$h_{ab} = h^+_{ba} \equiv h^*_{ab}, \tag{9}$$

---

[2] For one of the flavors f in Eq.(7), the boson operator $\varphi^{(C)} = i\sigma_y \varphi^+$ should be used as in SM [14].

(for isodoublets: $\Psi \to (\varphi^+\Psi)$).

The observed *CP*-violation, however, requires that the Yukawa coupling-matrices $h^f$ be complex-valued in the MS theory as well. The complex-valued $h$ are non-invariant to the time reversal *T* under *P* conservation. They certainly do not preserve *C*-symmetry either, but this results from the *CPT*-Theorem according to which it is impossible to write out a different local Lagrangian with desired properties. Therefore, the MS-Lagrangian is to be nonsymmetrical with respect to the "forward-backward" operation and, under certain conditions [17], can define time direction.

As mentioned earlier, only those transformations are acceptable in the generation space of the MS-theory that preserve condition (6). Therefore, it is possible to simultaneously bring two Yukawa couplings (7) for $f = \bar{u}, \bar{d}$ to a diagonal form even before MS breaking, since diagonalizing $T = \frac{1}{2}$ matrices are equal to each other: $U_{LR}^{\bar{u}} \equiv U_{LR}^{\bar{d}}$ (unlike SM). If *A* and $B^{(f)}$ are diagonal, the matrices $h^f$ are not arbitrary. Their most general form, with the left matrix $U_{LR}^+$ being the same, is:

$$h_{aa'}^{(\bar{u})} = \left(U_{LR}^+ h_{diag}^{(\bar{u})} U_{RL}^{(\bar{u})}\right)_{aa'}, \quad h_{aa'}^{(\bar{d})} = \left(U_{LR}^+ h_{diag}^{(\bar{d})} U_{RL}^{(\bar{d})}\right)_{aa'}, \tag{10}$$

Diagonalizing (7), which is necessary to obtain Dirac masses of mirror fermions (see Sections 3 and 4), gives:

$$A_{aa'} = \left(U_{LR}^+ A_{diag} U_{LR}\right)_{aa'}; \quad B_{bb'}^{(f)} = \left(U_{RL}^+ B_{diag} U_{RL}\right)_{bb'}^{(f)}. \tag{11}$$

This is the most general form of the mass terms in the MS-scheme. We have three directions in generation space: one direction for the distribution of diagonal Yukawa constants and the other two, for $\bar{u}, \bar{d}$-masses (4) of all MS-states. Changes between these directions are defined by the unitary matrices *U*, they do not break *SU*(2)-symmetry, and they do not affect the form of the Lagrangian's generation-diagonal terms, including the weak coupling (3). No mixing matrix appears here. In SM, analog diagonalization (7) is considered to involve four unitary matrices and it takes place only upon *SU*(2) breaking.

## 3.   Mirror Symmetry Spontaneous Breaking. CP-Properties of the Quark System

Let us build the Lagrangian for the isodoublet scalars $\phi_1$ and $\phi_2$ (8) which will provide the existence of two symmetrical minima:

$$\begin{aligned}\langle\phi_1\rangle &= \eta \neq 0, & \langle\phi_2\rangle &= 0; \\ \langle\phi_1\rangle &= 0, & \langle\phi_2\rangle &= \eta.\end{aligned} \tag{12}$$

Here, by $\langle\phi_1\rangle$ and $\langle\phi_2\rangle$ we mean the neutral component of the isodoublet.

It is possible to find the matching Lagrangian of self-action $\phi_1$ and $\phi_2$ by taking

$$V(\phi_1, \phi_2) = \kappa |\phi_1|^2 |\phi_2|^2 + v(\phi_1, \phi_2), \qquad (13)$$

as the potential, where the symmetrical expression $v(\phi_1 \phi_2)$ can be chosen, for example, to be the simplest potential for the Higgs boson in SM [14]:

$$v(\phi_1, \phi_2) = -\frac{\rho^2}{2}|\phi_1|^2 + \frac{\lambda}{4}|\phi_1|^4 + (\phi_1 \to \phi_2). \qquad (14)$$

From (13) and (14) we obtain two minima (12) with the standard vacuum mean value $\eta^2 = \rho^2/\lambda$.

The spontaneous breaking

$$\phi_1 = \frac{\eta + H}{\sqrt{2}} e^{i\sigma_k \theta_k}, \quad \phi_2 = \begin{vmatrix} \xi_1^+ + i\xi_2^+ \\ \xi_1^0 + i\xi_2^0 \end{vmatrix}, \qquad (15)$$

where $H$ is the Higgs boson, $\Theta_k$ is Goldstone variables providing masses of W-bosons, leads to mass matrices of the light particles $\psi^f$ having the form:

$$m_{ab}^{(f)} \simeq \sum_{n=0}^{2} A_{an} \frac{1}{\mu_n^{(f)}} B_{nb}^{(f)}, \quad \mu_n = h_n \eta. \qquad (16)$$

Here $n$ = 0, 1, 2 are numbers of the states $(\Psi_R \Psi_L)$, diagonalizing matrices $h^f$ in (7). The numbers $n$ correspond to the orders of the mass hierarchy (further see (18)). These are heavy mirror quarks with $\mu_n^{(f)}$ masses. The operator $\phi_2$ represents four, generally speaking, heavy bosons. They form two isodoublets, analogs of $K\overline{K}$-mesons (see [3]). The separable form (16) produces WMM with all qualitative properties of the observed CKM-matrix [1].

A number of questions arise with regard to Eq.(15) that defines *H* as the Higgs boson [15]. These questions are also discussed in [3] and in Section 5 of this paper. For instance, $\eta$ defines the mass of W-bosons and therefore $\eta = 246$ GeV [14]. According to Eqs.(7) and (15), the Higgs scalar *H* appears to interact with only heavy mirror particles $\Psi$. Large masses of $\Psi$ require non-perturbative Yukawa constants *h*. Interactions of *H* with the $\psi$ fermions of SM will inevitably appear as a direct consequence of the broken *SU*(2) symmetry. Their constants have a normal value of $m_{SM}/\eta$. This occurs owing to the small corrections $m_{SM}/M_{mir}$ in the eigenvalues of the mirror states $n$ = 0, 1, 2 (see [3] and Eq. (A.9) of Appendix). Changes in the *H* boson production mechanism associated with heavy mirror fermion participation were also discussed in [3].

The corrections are not taken into consideration in Eq.(16). Eq.(16) is an approximate solution to the mass matrix problem of SM's $\psi$ particles in the MS-theory, Sections 2 and 3. Eq.(16) is valid for

$$\mu \gg \left|\frac{AB}{\mu}\right| \sim m_{\text{SM}}. \tag{17}$$

A possibility of more accurate calculations is discussed in Appendix. For general properties included in the proposed MS-scenario, the corrections are of no interest.

For arbitrary complex-valued matrices *A* and *B* and the mirror mass hierarchy

$$\mu_0^{(f)} \gg \mu_1^{(f)} \gg \mu_2^{(f)} \tag{18}$$

Eq.(16) corresponds to a WMM with properties similar to the Wolfenstein matrix [10], and MS quark mass spectrum with hierarchy [1] inverse to (18).

Complexities of the unitary matrices *U* allow introduction of *CP*-violation into the MS-scheme. The properties of WMM for quarks remain here the same as in SM. The restrictions (10)-(11) are of no consequence. WMM can be expressed by the parameters of $A_n, B_n$ vectors and $\mu_n$ masses [1] or reduced to CKM standard parametrization [9].

## 4. Spontaneous Breaking of Mirror Symmetry. *CP*-Properties of the Lepton System

In the lepton WMM, there is no such remarkable phenomenon as the hierarchy of CKM matrix elements. A qualitatively different WMM can be reproduced in a number of ways. In [2], we chose a mechanism that is based on *SU*(2)-symmetry—that is, condition (6)—and simultaneously requires only one source of *P*-parity violation—MS breaking. The choice was purely phenomenological: its advantage is naturalness and consistency in reproduction of observed qualities. This mechanism is attractive due to a number of positive features and results:

1. It is similar to the mechanism of a quark system.
2. The choice of neutrino properties is unambiguous and clear—that is, Dirac neutrinos with inverse mass hierarchy.
3. A WMM with suitable properties is easy to build.
4. Formulae indicate that neutrino masses differ immensely from the masses of charged leptons, compared to a similar difference between masses of quark $\bar{u}$- and $\bar{d}$-families, at $M \gg \mu$ for the Majorana (*M*) and Dirac masses (*μ*) [13].

Therefore, as in [2], we assume that in addition to Yukawa couplings (7) and MS-lepton masses (4) and (5), *SU*(2)-invariant Majorana forms are also present for a neutrino flavor:

$$\left(\Psi_{RL}^T h_M C \Psi_{RL}\right)^{(\nu)} \varphi' + \left(\Psi_{RL}^T h_M C \gamma_5 \Psi_{RL}\right)^{(\nu)} \bar{\varphi}' + \text{c.c.}. \tag{19}$$

The $h_M^{(\nu)}$ matrices are identical in both terms of the sum (19) as was the case in Eq.(7). The isoscalar bosons $\varphi'$ and $\bar{\varphi}'$ acquire vacuum averages following a procedure similar to that in Section 3, i.e., by means of the combinations:

$$\phi = \varphi' - \bar{\varphi}' \qquad \text{and} \qquad \phi' = \varphi' + \bar{\varphi}'. \tag{20}$$

At the same time, the same scalar isodoublets $\varphi_1$ and $\varphi_2$ that defined quark couplings should be used in the lepton analog of the Yukawa couplings (7). Their vacuum averages form *W*-boson masses, and introduction of new isodoublets is considered an unnecessary complication.

Creating expected properties, i.e., phenomenology, requires that the Majorana mass also appear for the *R*-component of the isodoublet $\Psi_{LR}^{(\nu)}$: $\Psi_R = \frac{1}{2}(1 + \gamma_5)\Psi_{LR}$ [2]. In this case, the mass $M_R$ in this part of the Lagrangian must be with a minus sign equal to the mass $M_L$, which results from the spontaneous breaking in (19):

$$M_R = -M_L. \tag{21}$$

Condition (21) does not entail parity non-conservation in addition to MS-violation, and results in properties 2-4. The coupling (21) is what defines the Dirac nature of neutrino: two Majorana masses with the same absolute values.

In [3], a qualitative dynamic mechanism is described which is able to provide production of the Majorana mass $M_R$ in MS-breaking. This mechanism is intimately associated with and results directly from the interactions (2) and (19) and can support the appearance of the relationship in (21). Its non-perturbative character hinders analytical confirmation. On the other hand, scenarios commonly used to introduce Majorana mass terms for isodoublet fermions—that is, consideration of effective non-renormalizable Lagrangians with the square of the Higgs scalar $\phi_1^2$ or use of the isovector boson $\vec{\phi}$ [13]—present mechanisms so different from (19) that the existence in them of condition (21) seems absolutely improbable.

In the case of leptons, simultaneous diagonalization of the Yukawa terms (7) and interaction (19) must be possible also prior to MS-breaking. For quarks, this condition was imposed by the requirement that Eq.(6) be preserved in matrix form, which resulted in the observed hierarchy of CKM matrix elements. For the lepton WMM, such hierarchy is not observed, however, the same spectral pattern indicates a similarity between the mechanisms defining Dirac quark and charged lepton masses and, probably, the Dirac component of neutrino masses. We therefore assume that condition (6) is also valid for the lepton matrices *A*, and that:

$$U_{LR}^{(\nu)} = U_{LR}^{(e)} = U^{(\nu)}. \tag{22}$$

Then, the $h^{(v)}$ and $h_M^{(v)}$ matrices define the interaction of the isoscalar $\Psi_{RL}^{(v)}$, which upon diagonalization (7) and (19) is transformed by means of the same matrix $U_{RL}^{(v)}$. Furthermore, the parameters $h^{(v)}$ and $h_M^{(v)}$ collectively define the properties of the same states—Dirac neutrinos ([2], Section 6). The $U_{RL}^{(v)}$ matrix is symmetrical, and $U^{(v)}$ and $U_{RL}^{(e)}$ are unitary matrices.

If breaking of discrete symmetries entails only MS breaking, then Eqs.(7) and (19) are invariant to *T*-, and as a consequence, to *C*-transformations. Let us consider the C-invariance of the MS-Lagrangian lepton part. All interactions in this case involve only real-valued coupling matrices. For the Majorana terms (19), these matrices are not only real-valued, but also symmetrical:

$$\Psi_a^{T(c)} h_{ab} C \Psi_b^{(c)} = \bar{\Psi}_a C^T h_{ab} C \cdot C \bar{\Psi}_b^T = \bar{\Psi}_b h_{ab} C \bar{\Psi}_a^T, \qquad (23)$$
$$h_{ab} = h_{ba}^+ = h_{ab}^*.$$

Here, the "Majorana phases" [9] are absent. Diagonalization of Eq.(19) involves the orthogonal matrix *O*:

$$h_M^{(\nu)} = O^T \left(h_M^{(\nu)}\right)_{diag} O, \qquad (24)$$

For lepton systems, the only possibility to obtain *CP*-breaking phases in WMM is to introduce direct *T*-symmetry violation into the MS-Lagrangian. Then, the general form of the quantities containing and defining the complex properties of the theory is as follows:

$$A = U^{(\nu)+} A_{diag} U^{(\nu)}; \quad B^{(\nu)} = O_{RL}^{(\nu)+} B_{diag}^{(\nu)} O_{RL}^{(\nu)}; \quad B^{(e)} = U_{RL}^{(e)+} B_{diag}^{(e)} U_{RL}^{(e)}. \qquad (25)$$

Here, $U^{(v)}$ and $U_{RL}^{(e)}$ are arbitrary unitary matrices, $O_{RL}^{(v)}$ is an orthogonal, complex matrix. By means of these quantities, the complex factors pass into mass matrices (16) and WMM for SM leptons, including heavy mirror particles. Despite the Dirac nature of neutrinos, Majorana phases may also be present in the lepton WMM of the proposed scenario [9].

In addition to the violation of the total lepton number (see Section 6), such a situation creates possibilities for leptogenesis in MS-scenario.

## 5. General Properties of the Lagrangian with Broken Mirror Symmetry

The hypothetical Lagrangian based on the principle of MS and its spontaneous breaking consists of the following parts (omitting SM-terms):

$$\mathcal{L}_Y = \mu_n^{(f)} \bar{\Psi}_{Rn}^{(f)} \Psi_{Ln}^{(f)} + h_n^{(f)} \bar{\Psi}_{Rn}^{(f)} \Psi_{Ln}^{(f)} H + h_a^{(f)} (\bar{\psi}_L \phi_2)_a \psi_{Ra}^{(f)} + \text{c. c.}, \qquad (26)$$

$$\begin{aligned}\mathcal{L}_M = &\ A_{an}\left(\bar{\Psi}_{Rn}^{(f)}\psi_{La}^{(f)}\right) + B_{bn}^{(f)}\left(\bar{\Psi}_{Ln}^{(f)}\psi_{Rb}^{(f)}\right) + \text{c.c.}\ ; \\ & f = \bar{u}, \bar{d}, \quad a,b = 1,2,3, \quad n = 0,1,2.\end{aligned} \qquad (27)$$

Eqs.(26) and (27) are written out in terms of diagonal Yukawa couplings. The matrices *A* and *B* are Hermitian, Eqs.(11), (25). The constants $\mu$ and $h$ are real-valued.

For leptons, these expressions also include terms of Majorana type ($\bar{u} = v, \bar{d} = e$):

$$M_n \Psi_{Ln}^{(\nu)T} C \Psi_{Ln}^{(\nu)} + h_{M_n}^{(\nu)} \Psi_{Ln}^{(\nu)T} C \Psi_{Ln}^{(\nu)} \Phi + h_{M_b}^{(\nu)} \psi_{Rb}^{(\nu)T} C \psi_{Rb}^{(\nu)} \Phi' + \text{c. c.} \qquad (28)$$

and couplings appearing in one way or another ([3] and Section 4):

$$-M_n \Psi_{Rn}^{(\nu)T} C \Psi_{Rn}^{(\nu)} - h_{M_n}^{(\nu)} \Psi_{Rn}^{(\nu)T} C \Psi_{Rn}^{(\nu)} \Phi - h_{Ma}^{(\nu)} \psi_{La}^{(\nu)} C \psi_{La}^{(\nu)} \Phi' + \text{c.c.} \qquad (29)$$

The necessity in the first term in (29) and its sign are determined by phenomenology, i.e., selection of conditions that fit best the task of WMM reproduction and smallness of neutrino masses [2]. Other terms in (29) have no phenomenological support, however, they can result from the dynamic scenario in [3]. Their attractive feature is that, together with the contributions (28), they conserve parity, restricting the *P*-violation of the MS-system to weak interactions only. At any values of $A^{(\bar{u})} = A^{(\bar{d})}$ and $B^{(f)}$, the mass hierarchy $\mu_n$ and opposite signs in (28) and (29), these expressions result in the WMM structure for quarks and leptons imitating the observed properties.

Let us summarize some general effects of the MS-nature of these expressions. In part, they were previously discussed in [3].

1. If *H* is the real Higgs boson [15], then the Yukawa constants $h$ are large and non-perturbative. Indeed, then $\eta = 246$ GeV and $\mu_n = h_n \eta \gg m_{SM}$, which makes $h_n$ large. With the Yukawa constants being non-perturbative, the role of the Lagrangians (26) to (29) is limited strictly to representation of the MS principles.

2. In Eq.(26), the *H*-boson interacts with the $\Psi$ components only. On the other hand, the interaction of *H* with $\psi$ particles—that is, with SM fermions—is present as a result of (26) and (27). The constants of this interaction, which is diagonal in terms of generation indices, are perturbative, since they coincide with the standard constants $m_{SM}/\eta$. The weakening mechanism is related with the transitions $\Psi \leftrightarrow \psi$ (27), and its properties, with the gauge symmetry *SU*(2) and corrections $\beta$ and $\delta$ in Eqs.(A9) for precise eigenfunctions. This question is discussed more thoroughly in [3].

3. The diagonalization of the mass terms in (26) and (27) for $\Psi$ and $\psi$ particles does not affect the diagonalization of flavor-neutral weak and other interactions in terms of generation indices. This is also true for the interaction of the *H*-boson with the particles of SM. In lower orders of $m_{SM}/M_{mir}$, this is easy to verify, based on unitary transformation matrices and symmetry consequences. Further approximations (see Appendix) depend on the structure and concrete values of the parameters of the matrices *h*, *A*, *B*, *μ*, and *M*. Conditions for the numerical values and quantities of unknown masses $M_{mir}$ impose restrictions on these small, order-of-magnitude terms $(m_{SM}/M_{mir})^K$.

4. Owing to the transitions $\Psi \leftrightarrow \psi$ (27), all heavy mirror particles become unstable. They decay into particles of SM. This can also occur through weak interactions, with $m_{SM}/M_{mir}$ being an additional smallness. The mechanism is again related with minor corrections for the eigenvalues of physical states (see Eq.(A9)).

## 6.    Interactions of Mirror Neutrinos

Weak interactions of mirror neutrinos and their relationship with the Higgs boson *H* do not preserve the total lepton number. This is also a general consequence of the Lagrangians in Section 5.

The expression for the mass terms of mirror neutrinos is:

$$\mu_n^{(\nu)} \bar{\Psi}_{Rn}^{(\nu)} \Psi_{Ln}^{(\nu)} + M_n \left( \Psi_{Ln}^{(\nu)T} C \Psi_{Ln}^{(\nu)} - \Psi_{Rn}^{(\nu)T} C \Psi_{Rn}^{(\nu)} \right) + \text{c.c.} \tag{30}$$

We ignore the transitions $\Psi \leftrightarrow \psi$ (27); their consideration would be the approximation $m_{SM}/M_{mir}$. Eq.(30) can be rewritten in terms of Majorana operators (we repeat here [2]) omitting the symbol *ν*:

$$\tilde{\Psi}_R = \frac{\Psi_R + C\bar{\Psi}_R^T}{\sqrt{2}}, \quad \tilde{\Psi}_L = \frac{\Psi_L + C\bar{\Psi}_L^T}{\sqrt{2}}. \tag{31}$$

Then Eq. (30) becomes

$$\mu_n \left( \bar{\tilde{\Psi}}_{Rn} \tilde{\Psi}_{Ln} + \bar{\tilde{\Psi}}_{Ln} \tilde{\Psi}_{Rn} \right) + M_n \left( \bar{\tilde{\Psi}}_{Rn} \tilde{\Psi}_{Rn} - \bar{\tilde{\Psi}}_{Ln} \tilde{\Psi}_{Ln} \right), \tag{32}$$

which corresponds to two Majorana particles with the same mass but opposite signs:

$$\lambda_\pm = \pm\sqrt{M^2 + \mu^2} = \pm\lambda. \tag{33}$$

For each *n*, eigenfunctions of these states are:

$$X_+ = \frac{1}{N}\left( \tilde{\Psi}_R + \frac{\mu}{M+\lambda} \tilde{\Psi}_L \right), \quad X_- = \frac{1}{N}\left( \tilde{\Psi}_L - \frac{\mu}{M+\lambda} \tilde{\Psi}_R \right), \tag{34}$$

where *N* is a normalization coefficient. These eigenfunctions represent a single Dirac spinor with the mass *λ*:

$$X = \frac{X_+ + \gamma_5 X_-}{\sqrt{2}}, \quad X^C = \frac{X_+ - \gamma_5 X_-}{\sqrt{2}}, \tag{35}$$

$X^C = C\bar{X}^T$ is the charge conjugate spinor.

From Eqs.(34) and (35), we have:

$$\begin{aligned}\tilde{\Psi}_R &= \frac{1}{\sqrt{2}\,N}\left\{\left(1 - \frac{\mu}{M+\lambda}\gamma_5\right)X + \left(1 + \frac{\mu}{M+\lambda}\gamma_5\right)X^C\right\}, \\ \tilde{\Psi}_L &= \frac{1}{\sqrt{2}\,N}\left\{\left(\gamma_5 + \frac{\mu}{M+\lambda}\right)X - \left(\gamma_5 - \frac{\mu}{M+\lambda}\right)X^C\right\}.\end{aligned} \quad (36)$$

Substituting these formulae in the weak current $\Psi_R$, Eq.(3), for interactions with charged $W$ we obtain:

$$\begin{aligned}gW_\mu\bar{\Psi}_R^{(e)}\gamma^\mu\Psi_R^{(\nu)} &\equiv \sqrt{2}gW_\mu\bar{\Psi}_R^{(e)}\gamma^\mu\tilde{\Psi}_R^{(\nu)} = \frac{g}{N}W_\mu\bar{\Psi}_R^{(e)}\gamma^\mu \times \\ &\times \left[\left(1 - \frac{\mu}{M+\lambda}\gamma_5\right)X_R + \left(1 + \frac{\mu}{M+\lambda}\gamma_5\right)C\bar{X}_L^T\right],\end{aligned} \quad (37)$$

and for neutral current, we have:

$$\begin{aligned}gW_\mu^{(0)}\bar{\Psi}_R^{(\nu)}\gamma^\mu\Psi_R^{(\nu)} &= \frac{g}{N^2}W_\mu^{(0)}\Big\{2\bar{X}\left(1 + \frac{\mu}{M+\lambda}\gamma_5\right)^2\gamma^\mu\gamma_5 X + \\ &+ \left(1 - \frac{\mu^2}{(M+\lambda)^2}\right)\left[X^T C\gamma^\mu\gamma_5 X + \bar{X}\gamma^\mu\gamma_5 C\bar{X}^T\right]\Big\}.\end{aligned} \quad (38)$$

In (37) and (38), we have non-conservation of the total lepton number. The weak interaction of charged leptons and electromagnetic interaction remain unchanged from the usual form.

A similar situation is observed for the interaction of mirror neutrinos (37) with the Higgs boson $H$:

$$\begin{aligned}h^{(\nu)}H(\bar{\Psi}_R\Psi_L + \bar{\Psi}_L\Psi_R) &\equiv h^{(\nu)}H\left(\bar{\tilde{\Psi}}_R\tilde{\Psi}_L + \bar{\tilde{\Psi}}_L\tilde{\Psi}_R\right) = \\ &= \frac{h^{(\nu)}H}{N^2}\left\{\frac{4\mu}{M+\lambda}\bar{X}X - \left(1 - \frac{\mu^2}{(M+\lambda)^2}\right)\left(\bar{X}\gamma_5 C\bar{X}^T - X^T C\gamma_5 X\right)\right\}.\end{aligned} \quad (39)$$

Eqs.(38) and (39) are related to each other by $SU(2)$-gauge invariance violation. Since for the real Higgs boson the constant $h^{(\nu)}$ must be non-perturbatively large, Eq. (39) has a purely symbolic sense: lepton number non-conservation in the "mirror world" results not only from weak interactions.

Similar phenomena are also observed for interactions with $\Phi, \Phi'$, Eqs. (28) and (29).

## 7. Conclusion

Results of *CP*-violation lepton phase measurements show mild preference for the normal hierarchy of neutrino masses and the large phase value ([19] and proceedings of the CERN seminar of 17.12.25). The normal hierarchy does not meet the expectations of the MS-breaking model proposed in the present paper. According to the authors of paper [19], their results are very preliminary; since only a small portion of planned statistics are available to date. Papers [20,21] also point to insufficient reliability of the results.

In the MS-violation models being discussed, a second Higgs boson must be present—that is, two neutral (CP = ±1) and two charged scalars (see also [22]). Their masses are unknown and may be very large. Any assessments as to their origins and decays and influence on other processes are complicated due to their non-perturbative couplings with SM fermions. Yet, the second heavy isospinor scalar is a mandatory element in MS-breaking.

Let us mention a number of MS-scheme features that were previously discussed in [2,3]. Neutral transitions between generations could exist here in $\sim m_{SM}/M_{mir}$ and smaller corrections. Experimental limitations [18] can specify the values of mirror state masses. Any quantitative evaluations are again impossible due to the presence of non-perturbative Yukawa couplings. Such processes, however, are not MS specific—they are typical of practically any scenario outside SM. Mirror fermions that could be the lightest in mass correspond to the heaviest masses of SM, namely, $t$-quarks and $\tau$-leptons. This is a consequence of inverse hierarchy (16) for SM masses and mirror particles. Inverse hierarchy increases the coupling constants $h$ exactly for fermions with small masses. Since the constants are non-perturbative, their characteristics are uncertain.

Thus, a number of important, yet unexplained phenomenological properties can be interpreted by means of broken MS at very high energies. These properties include the structure of mixing matrices, the difference of mixing patterns for quarks and leptons, and a particular smallness of neutrino masses. This paper shows that the leptogenesis mechanism can also be included in this interpretation.

At the same time, MS-interpretation requires that mirror fermion masses be very heavy. The inverse coupling of "normal" and mirror masses, (16) and (18), and an exceptional smallness of neutrino masses make one suggest that even the lightest of mirror particles have masses that are much larger than the mass of $t$-quark, $m_t \simeq 173$ GeV [9]. Thus, there are only indirect indications available so far to confirm our hypothesis.

The author is grateful to Ya. I. Azimov and M.G. Ryskin for their interest in this work and for useful discussions. This work was funded by grant RSF No. 14-92-0028.

## 8. Appendix

Let us explain the mechanism underlying the mass formula (16), its approximate character, and methods of calculating subsequent terms of expansion with respect to the parameter $m/\mu \sim$ $AB/\mu^2 \ll 1$ using the simple example of the well-known mass calculations [13] in the see-saw problem.

In terms of Majorana operators (31), the diagonalization of the mass matrix

$$M\left(\psi_R^T C \psi_R + \bar{\psi}_R C \bar{\psi}_R^T\right) + \mu\left(\bar{\psi}_R \psi_L + \bar{\psi}_L \psi_R\right) \tag{A.1}$$

is reduced to the matrix

$$\begin{array}{c} \bar{\tilde{\psi}}_R \;\; \bar{\tilde{\psi}}_L \\ \left|\begin{array}{cc} M & \mu \\ \mu & 0 \end{array}\right| \begin{array}{c} \tilde{\psi}_R \\ \tilde{\psi}_L \end{array} \end{array} \tag{A.2}$$

with the characteristic equation

$$\lambda^2 - M\lambda - \mu^2 = 0 \tag{A.3}$$

and the eigenvalues

$$\lambda_\pm = \frac{M}{2} \pm \sqrt{\frac{M^2}{4} + \mu^2} \simeq \begin{cases} M + \mu^2/M \\ -\mu^2/M \end{cases}. \tag{A.4}$$

The operation of diagonalization could be performed by the diagram method, using only the Lagrangian rewritten in terms of $\tilde{\psi}$ (see (32)). Let us denote masses of the two particles described by this Lagrangian as $m_1$ and $m_2$. For these masses, we then obtain equations corresponding to the Feynman pole diagrams in Fig.1:

$$m_1 = -\mu \frac{1}{M - m_1} \mu, \quad m_2 = M - \mu \frac{1}{-m_2} \mu. \tag{A.5}$$

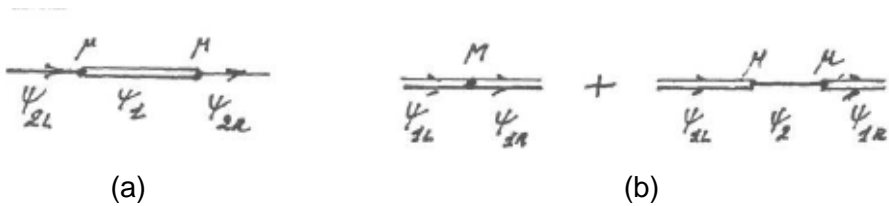

(a)            (b)

Fig.1 Diagrams for see-saw mass calculation: $\lambda_-$ (a); $\lambda_+$ (b)

In the first equation, the propagator momentum is $\hat{p} \equiv m_1$. In the second, we have $\hat{p} \equiv m_2$. Both equations (A.5) lead to the same characteristic equation (Eq.A.3). It is obvious that $\lambda_\pm$ should have opposite signs in the two equations (A.5). Subsequent terms of expansion with respect to $\mu \ll M$ are easy to determine.

In a problem with many generations, we can apply a similar procedure, where Eq.(16) will be its first term. The diagonalization problem for the mass operator

$$\mathcal{M}_{ab} = \sum_n A_{an} \frac{1}{\mu_n - \hat{p}} B_{nb}, \quad |\mu| \gg |A|, |B|. \tag{A.6}$$

can be solved in several steps:

- Bring $\widehat{\mathcal{M}}$ to a diagonal form with an arbitrary $\hat{p}$ ($\mathcal{M}$ is Hermitian here):

$$\text{diag}\left(\tilde{m}_1(\hat{p}), \tilde{m}_2(\hat{p}), \ldots, \tilde{m}_n(\hat{p})\right) = U^+(\hat{p})\widehat{\mathcal{M}}U(\hat{p}). \tag{A.7}$$

- Equate:

$$m_1 = \tilde{m}_1\left(\hat{p} = m_1\right), \ldots, \tilde{m}_n\left(\hat{p} = m_n\right), \tag{A.8}$$

and solve these equations to any precision.

- The eigenfunctions of the problem are the rows $U_n(\hat{p} = m_n)$, $U_{n'}(\hat{p} = m'_n)$…, which are orthogonal to one another. Same is true for states with large masses $\mu_n$.

Precise eigenstates of the complete problem have the form of superposition:

$$\begin{aligned} \Psi^{(n)} &= \alpha_{ni}\Psi_i + \beta_{ni}\psi_i, \\ \psi^{(n)} &= \gamma_{ni}\psi_i + \delta_{ni}\Psi_i, \end{aligned} \tag{A.9}$$

Where $\Psi_i, \psi_i$ are eigenfunctions of the problem (A.6) and its analog for $\Psi$ (similar to the two equations in Eq.(A.5)). The matrices $\beta$ and $\sigma$ in (A.9) are matrices with elements as small as, or smaller than, $\sim \left(\frac{m}{\mu}\right)^{1/2}$. These additional terms define, among other things, the interaction of the Higgs boson $H$ with fermions of SM—that is, the smallness of the Higgs Yukawa constants $m_{SM}/\eta$ (Section 5).